\documentclass[12pt,journal,compsoc]{IEEEtran}
\usepackage{lineno}
\usepackage{graphicx}

\usepackage{scalefnt}
\usepackage{url}

\usepackage{tikz}
\usetikzlibrary{shapes,arrows}

\usepackage[ruled,vlined,linesnumbered]{algorithm2e}

\ifCLASSOPTIONcompsoc
\else
  \usepackage{cite}
\fi
\ifCLASSINFOpdf
\else
\fi
\begin{document}
%
\title{Wireless Message Dissemination via Selective Relay over Bluetooth (MDSRoB)}
%
%
%

\author{Joseph Paul Cohen\\
joecohen@ieee.org
\IEEEcompsocitemizethanks{\IEEEcompsocthanksitem Joseph Paul Cohen is with the
Department of Computer Science at The University of Massachusetts Boston, Boston, Massachusetts, 02125.\protect\\
E-mail: joecohen@ieee.org}
}
\IEEEcompsoctitleabstractindextext{%
\begin{abstract}
This paper presents a wireless message dissemination method designed with no
need to trust other users.  This method utilizes modern wireless adaptors
ability to broadcast device name and identification information.  Using the scanning
features built into Bluetooth and Wifi, messages can be exchanged via their
device names. This paper outlines a method of interchanging multiple messages to
discoverable and nondiscoverable devices using a user defined scanning interval 
method along with a response based system.  By selectively relaying messages
each user is in control of their involvement in the ad-hoc network.

\end{abstract}


}

\maketitle

\IEEEdisplaynotcompsoctitleabstractindextext

%
\IEEEpeerreviewmaketitle

\section{Introduction}

In disintermediation methods between wireless devices there are many challenges
to compatibility in devices.  Software API's were not designed for this until
the most recent versions of Bluetooth stacks. Also Bluetooth software stacks
contain unique bugs which make it hard to ensure that all features work across
devices. Wireless adapter APIs on consumer cell phones have not allowed
communication directly between devices until the most recent versions. Bluetooth
has allowed direct communication but requires pairing.

Pairing Bluetooth devices brings with it a security risk because a paired device
can access all exposed services on a device.
Pairing provides a false sense of security in that the user expects their
Bluetooth headphones to only be able to receive music. Pairing with a Bluetooth
headset will also allow the headset access to phonebooks and other Bluetooth
services.

The goal of this project is to design an communication method that works across
as many Bluetooth implementations as possible. And also to achieve this
communication without pairing or any previous interaction. As well as no special
configuration of the device such as special drivers or root access.

\section{Related Work}

\begin{figure}[htp]
\begin{center}
  \includegraphics[width=0.4\textwidth]{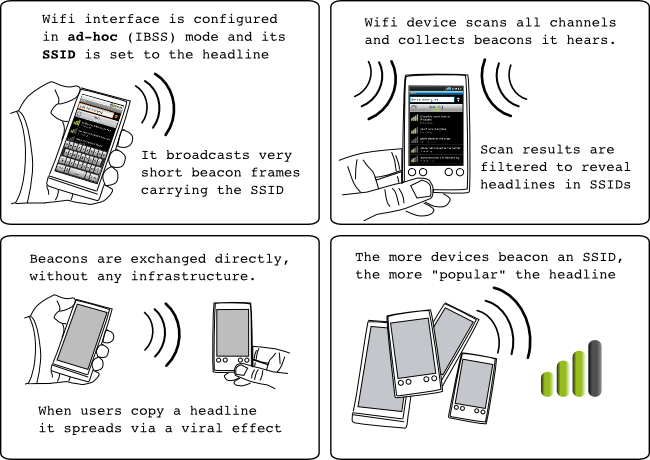}
  \caption[labelInTOC]{Dythr Project \cite{jakubczak_dythr_2010}}
  \label{fig:dythr}
\end{center}
\end{figure}

The most closely related project is from 2010 called Dythr
\cite{jakubczak_dythr_2010} and uses a method where a phone broadcasts a wifi
hotspot with the SSID being the message.  This method however requires root
privileges on the target android phones and cannot be used by the general
population.  It also requires support for each devices network card which lowers
the utility even more. The main use case of this work is very similar to the
proposed method and a graphic from their project is shown in Figure \ref{fig:dythr}.

In 2010 Huang et al. \cite{huang_phonenet:_2010} proposed PhoneNet. This method
uses a central server to establish links between devices connected to a wifi
network and then allows devices connected on local networks to connect directly.  

More recently; researchers at Standford have worked on disintermediation of
social networks. Dodson et al. \cite{purtell_musubi:_????} proposed SocialKit
which allows app developers to utilize social networks without devoting to a
single service. This paper mentions wireless P2P networking but then does not
clearly explain the methods used.

Stanford has a lab called Mobisocial \cite{mobisocial_stanford_2012} which is
moving in the same direction as the ideas presented in this paper but I was
unable to find a clear publication that is in the same realm as this work.

\section{Blucat}

Typical Bluetooth usage requires a paired connection to directly connect with
other devices in order to establish communication between devices. To explore
the extent which Bluetooth can operate without pairing a tool was created called
Blucat \footnote{\url{http://blucat.sourceforge.net/}} that exercises the Java
Bluecove API. Blucat is based off the bluecove libraries and is designed to work
on Linux, Mac, and other systems.

Bluetooth offers a Service Discovery Protocol (SDP) for unpaired/unestablished
communication.  Over this protocol a set a service records can be exchanged
without pairing. This is exposed as a L2CAP service listening on channel 1. To
explore this, Blucat was created to exercise the Java API's available via the
bluecove project.

Blucat is designed to act like
netcat\footnote{\url{http://netcat.sourceforge.net/}} and
ncat\footnote{\url{http://nmap.org/ncat/}} as well as and have scanning features
similar to nmap\footnote{\url{http://nmap.org/}}.
There are many protocols inside Bluetooth instead of just TCP and UDP which
leads to some tough design decisions.

Blucat's nmap like ability to scan and discover devices in a piconet currently
uses SDP as well as brute force scan on both RFCOMM and L2CAP channels.

To examine the Service Discovery Method in Bluetooth stacks, Blucat performs
a General/Unlimited Inquiry Access Code (GIAC) discovery and returns the devices
found. The character limit observed for the devices tested was 248 characters
long. Bluetooth stacks such as Apple OSX, GPL Bluez, and Android cache the
device name which introduces lag into the device name update cycle.

\begin{verbatim}
$blucat devices                                                                 
Searching for devices
123456789000, "Nexus 7", ...
012345678900, "GT-P1010", ...
001234567890, "Android Dev Phone 1"
Found 3 device(s)
\end{verbatim}

Each device is queried for the RFCOMM UUID (0x0003) and the Service Name
attribute (0x0100). This offers another method of data transmission but the
Android Bluetooth stack does not support reading these names. Without Android
support this method of data transmission cannot be used.

{\scalefont{0.75}
\begin{verbatim}
$blucat services
Listing all services
Searching for services on 123456789000 Nexus 7
123456789000, "Nexus 7", "Test Service Name", ..
123456789000, "Nexus 7", "Hello world!!!", ..
123456789000, "Nexus 7", "OBEX Object Push", ..
Searching for services on 012345678900 GT-P1010
012345678900, "GT-P1010", "OPP Server", ..
012345678900, "GT-P1010", "FTP Server", ..
\end{verbatim}
}

For each service that corresponds to a RFCOMM channel, Blucat can establish a
socket and map $stdin$, $stdout$ and $stderr$ from the remote Bluetooth service
to the local command line.

Blucat sockets use RFCOMM by default because of it's goal to emulate serial
connections such as TCP sockets and RS-232. RFCOMM, also known as the Serial
Port Profile, is already used to interact with many devices such as headsets and
printers.

\section{MDSRoB}

Wireless Message Dissemination via Selective Relay over Bluetooth facilitates a
set of strings contained on each relay node to be received by other relay nodes.
One important restriction is that at least one relay node must have it's
Bluetooth adapter visible.  Modern versions of Android allow for indefinite
viability that will never timeout which make this restriction more reasonable.

The current string broadcast is basic and only includes a message with a
predefined header. 

$$\underbrace{JPC}_{header}message$$

Here I will detail a more useful version. Which includes a message id to make
referencing previous messages possible as well as compression using bzip2 and
base64 encoding. Also included is a one character type value to allow messages
to be encrypted using preshared keys to allow confidential dissemination of
messages. The symbol $|$ is reserved as a divider and if used in a message can
be escaped with $\backslash |$

$$\underbrace{MDSR}_{header}\overbrace{0}^{Type}\underbrace{id|message}_{compressed/encrypted}$$

\begin{table}[h]
\begin{tabular}{|c|c|}
\hline
0 & bzip2 and base64 \\
\hline

1 & bzip2 and RSA encryption and bzip2 and base64 \\
\hline
\end{tabular}
\caption{Message type values}
\label{tab:msgtypes}
\end{table}

What should also be specified in a more detailed description is a standard
substitution table for common English words that would be shipped with the
implementation of the protocol. This would allow a predefined optimal mapping
between common words and short string codes such as ``subway'' $\rightarrow$
``$\backslash$ sw''. Being able to consume space on each relay should increase
the possible message size. The symbol $\backslash$ is reserved as a escape
character and if used in a message can be escaped with $\backslash \backslash$

\begin{figure}[htp]
\begin{center}
  \includegraphics[width=0.5\textwidth]{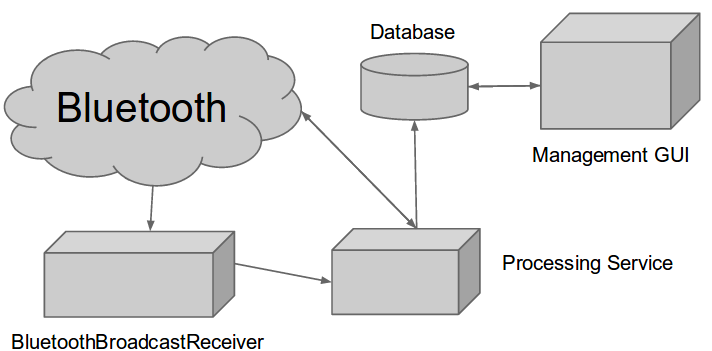}
  \caption[labelInTOC]{MDSRoB System Overview}
  \label{fig:system}
\end{center}
\end{figure}

The system overview is shown in Figure \ref{fig:system}. The
BluetoothBroadcastReceiver listens for device interaction with the Bluetooth
adapter and queues every device it can glean to be processed by the processing
service. The processing service then records the name of these devices, which
contain the message, and stores them in a database. It then proceeds to send
it's messages to the device by modifying it's name and contacting the remote
device so it's BluetoothBroadcastReceiver is triggered to store the message.

\begin{figure*}[htp]
\begin{center}
  \includegraphics[width=0.7\textwidth]{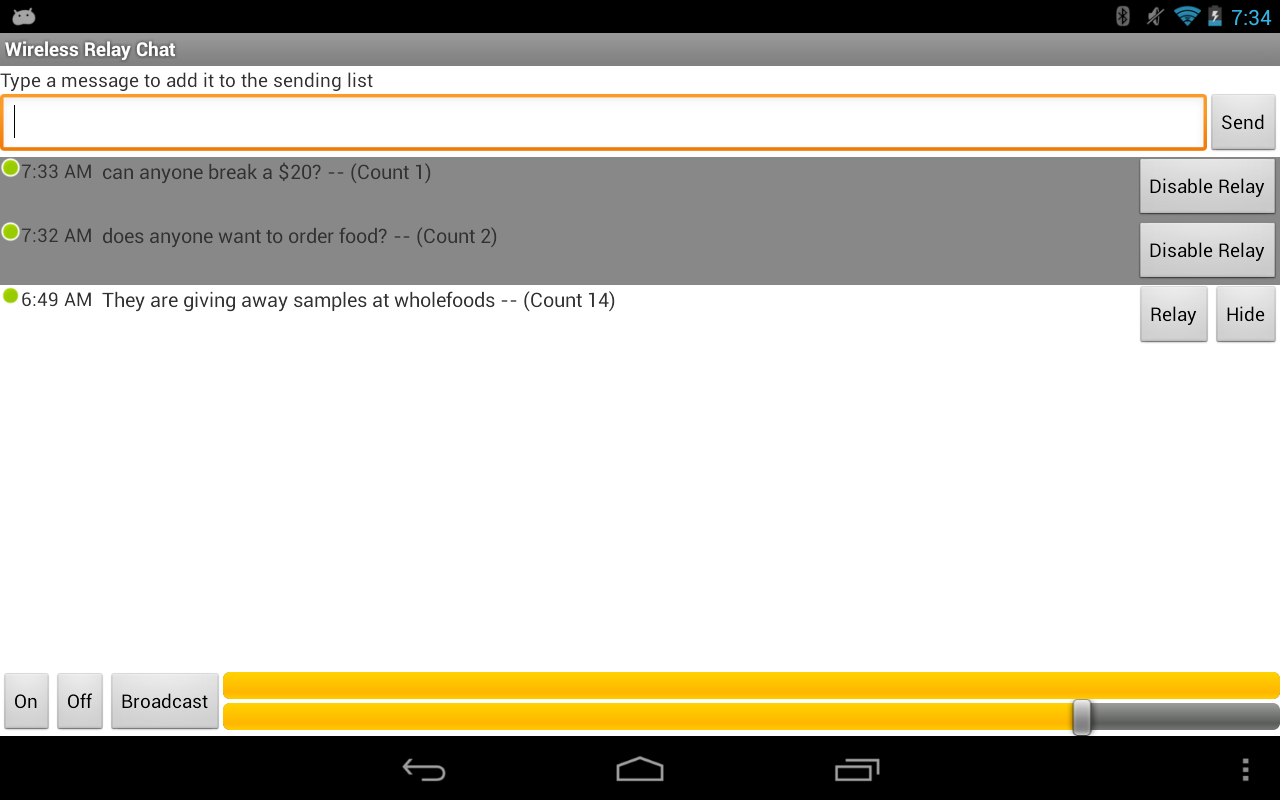}
  \includegraphics[width=0.7\textwidth]{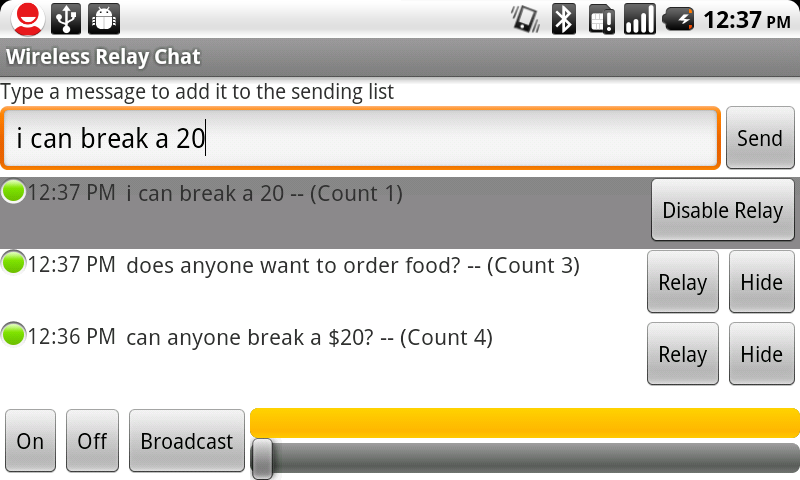}
  \includegraphics[width=0.7\textwidth]{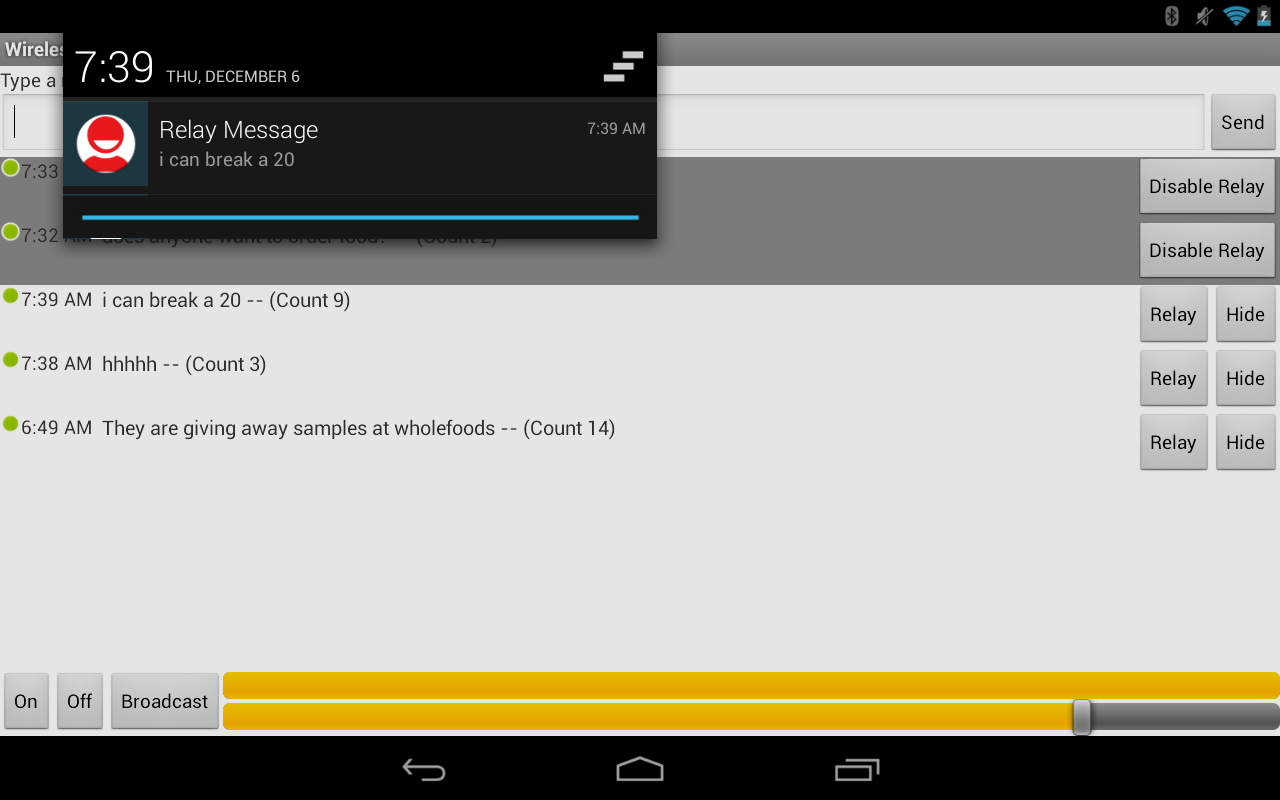}
  \caption[labelInTOC]{A sample interaction using the prototype.}
  \label{fig:sample}
\end{center}
\end{figure*}

The management GUI deals with displaying messages and setting which messages are
relayed or not. A sample interaction using the prototype GUI is shown in Figure
\ref{fig:sample}.First a device broadcasts two messages. The second user
responds by  broadcasting a response to everyone including the sender if they
are in the area. 

\begin{figure}[htp]
\begin{center}
  \includegraphics[width=0.5\textwidth]{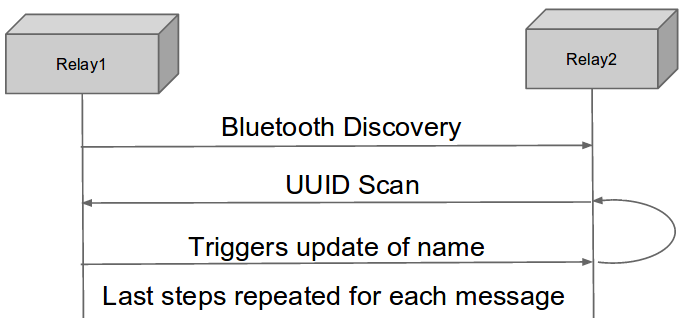}
  \caption[labelInTOC]{Relay node interaction}
  \label{fig:nodeinteraction}
\end{center}
\end{figure}

When a relay node contacts another relay node a set of steps occur. These are
shown in Figure \ref{fig:nodeinteraction}.  First a node will trigger the
process by preforming a Bluetooth discovery and causing a connection the remote
relay device.  Relay2 will then become aware of Relay1 and set its device name
to a message and perform a UUID scan to force Relay1 to update it's name. This
is repeated until all relayed messages are sent.

%

\ifCLASSOPTIONcaptionsoff
  \newpage
\fi



%

\bibliographystyle{IEEEtran}
\bibliography{joe,adhoc}

\end{document}